\documentclass[aps, prb, twocolumn, superscriptaddress, citeautoscript, floatfix, amsmath, amssymb]{revtex4-1}

\usepackage{lmodern,microtype}
\usepackage[mathlines]{lineno}
\relax

\usepackage{lipsum}
\usepackage{graphicx}
\usepackage{color,soul}
\usepackage[caption=false]{subfig}
\usepackage{array,booktabs,paralist,dcolumn}
\usepackage{enumitem}
\usepackage{multirow}
\usepackage{mathtools}
\usepackage{mathdesign,bm}
\usepackage[dvipsnames]{xcolor}

\usepackage{hyperref}
\usepackage{filecontents}

\setlist[itemize]{leftmargin=*}

\begin{document}

\date{\today}

\title{Polar magneto-optical Kerr effect in antiferromagnetic M$_2$As (M=Cr, Mn, Fe) under an external magnetic field}

\author{Kisung Kang}
\affiliation{Department of Materials Science and Engineering, University of Illinois at Urbana-Champaign, Urbana, IL 61801, USA}

\author{Kexin Yang}
\affiliation{Department of Physics, University of Illinois at Urbana-Champaign, Urbana, IL 61801, USA}

\author{Krithik Puthalath}
\affiliation{Department of Materials Science and Engineering, University of Illinois at Urbana-Champaign, Urbana, IL 61801, USA}
\affiliation{Department of Physics, University of Illinois at Urbana-Champaign, Urbana, IL 61801, USA}

\author{David G. Cahill}
\affiliation{Department of Materials Science and Engineering, University of Illinois at Urbana-Champaign, Urbana, IL 61801, USA}
\affiliation{Department of Physics, University of Illinois at Urbana-Champaign, Urbana, IL 61801, USA}
\affiliation{Materials Research Laboratory, University of Illinois at Urbana-Champaign, Urbana, IL 61801, USA}

\author{Andr\'e Schleife}
\email{schleife@illinois.edu}
\affiliation{Department of Materials Science and Engineering, University of Illinois at Urbana-Champaign, Urbana, IL 61801, USA}
\affiliation{Materials Research Laboratory, University of Illinois at Urbana-Champaign, Urbana, IL 61801, USA}
\affiliation{National Center for Supercomputing Applications, University of Illinois at Urbana-Champaign, Urbana, IL 61801, USA}

\begin{abstract}
Antiferromagnetic metals attract tremendous interest for memory applications due to their expected fast response dynamics in the terahertz frequency regime.
Reading from and writing information into these materials is not easily achievable using magnetic fields, due to weak high-order magneto-optical signals and robustness of the magnetic structure against external magnetic fields.
Polarized electromagnetic radiation is a promising alternative for probing their response, however, when ideal antiferromagnetic symmetry is present, this response vanishes.
Hence, in this work we combine first-principles simulations with measurements of the polar magneto-optical Kerr effect under external magnetic fields, to study magneto-optical response of antiferromagnetic M$_2$As (M=Cr, Mn, and Fe).
We devise a computational scheme to compute the magnetic susceptibility from total-energy changes using constraints on magnetic-moment tilting.
Our predictions of the spectral dependence of polar magneto-optical Kerr rotation and ellipticity allow us to attribute these effects to breaking of the magnetic symmetry.
We show that tilting affects the exchange interaction, while the spin-orbit interaction remains unaffected as the tilting angle changes.
Our work provides understanding of the polar magneto-optical Kerr effect on a band structure level and underscores the importance of the magnetic susceptibility when searching for materials with large magneto-optical response.
\end{abstract}

\maketitle

\section{\label{sxn:intro}Introduction}

As society becomes increasingly data driven, the ability to store and access enormous amounts of information quickly has become of utmost importance.
It keeps up with the growing global demands of high-performance electronics.
Random access memory \cite{Daughton:1992}, used in computing devices, have seen tremendous growth owing to giant magnetoresistance.
Spin valves have paved the way for the development of magnetic random access memory based on magneto-tunneling resistance \cite{Bhatti:2017}.
However, the magnitude of the anisotropy field in the ferromagnetic FM materials limits the response dynamics to the gigahertz range.

Antiferromagnetic (AF) materials (AFMs) are currently attracting tremendous interest \cite{Siddiqui:2020}, since they can possibly overcome some of these limitations:
Due to the strong exchange interaction between sublattice magnetic moments, AFMs are expected to have fast response dynamics in the terahertz frequency regime.
Additionally, unlike FM materials, AFMs exist in each magnetic symmetry group, providing a vast space of magnetic-moment configurations \cite{Baltz:2018}.
Hence, there is a large number of candidate AFMs and they are ubiquitous in the form of metals, semi-metals, insulators, and superconductors \cite{Baltz:2018}.
However, their characterization via magnetometer measurements is difficult due to the small difference between domains, whereas neutron diffraction requires large-scale facilities.

In addition, it has been shown that reading information from and writing into AFMs using magnetic fields is not easily achievable due to faint high-order magneto-optical signals and robust magnetic structure against external fields.
In contrast to FM materials, where the magnetic moments are aligned in a preferential easy axis direction, collinear AFMs exhibit alternating and non-collinear AFMs exhibit more complex geometries of magnetic moments throughout the lattice, resulting in vanishing net magnetization.
While this is the origin of the robustness of AFMs to external magnetic fields and allows them to withstand interference from deleterious stray fields, it also renders them invisible to magnetic probes.
Reading or writing of information in AFMs is difficult but achievable in metallic AFMs through electrical methods\cite{Wadley:2016, Meinert:2018}.

Polarized electromagnetic radiation is a promising alternative probe of material response \cite{Nemec:2018}, suitable for characterization of magnetic moment directions in the lab \cite{Oppeneer:2017}.
In particular, the magneto-optical Kerr effect (MOKE) \cite{John:1877} has been used to probe electronic and magnetic properties by measuring the polarization rotation of reflected light under various geometries.
From a response function point of view, linear MOKE for collinear antiferromagnets is related to the first-order term of an expansion of the dielectric tensor into increasing orders of net magnetization and N\'eel vector \cite{tzschaschel2017ultrafast,Iida2011,Eremenko2012,Yang2019,Siddiqui:2020}.
Oppeneer \textit{et al.}\ showed for FMs that non-vanishing linear MOKE requires presence of both spin-orbit coupling and exchange splitting in a material \cite{Oppeneer:1992}.
However, due to the vanishing net magnetization, majority and minority spin states are degenerate in collinear AFMs, hence, exchange splitting is absent.
In order to use linear MOKE to study collinear AFMs, one of the simplest approaches is to apply an external magnetic field to generate a small net magnetization along the field direction.
Linear MOKE can then be distinguished into polar, longitudinal, and transversal geometry, depending on whether the net magnetization is oriented out-of-plane to the surface, in-plane to the surface and in the plane of incidence, or in-plane to the surface and perpendicular to the plane of incidence, respectively.
In this work we focus on polar MOKE (PMOKE).

To this end, we perform first-principles simulations based on density functional theory (DFT) to study the ground and moment-tilted states of antiferromagnetic M$_2$As (M=Cr, Mn, and Fe).
Due to their high N\'eel temperatures above room temperature of 393 K \cite{Yamaguchi:1972}, 573 K \cite{Austin:1962}, and 353 K \cite{Katsuraki:1966}, respectively, these materials have potential for utilization in spintronic devices at room temperature.
Moment tilting is imposed in our first-principles approach to account for the effect of an external magnetic field acting on M$_2$As and we compute the magnetic susceptibility from the change in total energy.
We also perform first-principles simulations to investigate the PMOKE signal for light incidence parallel to the $c$-axis.
In addition, we implement PMOKE measurement on antiferromagnetic Fe$_2$As single crystals.
Finally, we discuss spin-orbit coupling and exchange splitting in the electronic band structure for the ground and spin-tilted state, to unveil the origin of Kerr rotation and ellipticity spectra.

\section{\label{sxn:comp}Computational Approach}

We perform fully relativistic, non-collinear first-principles density functional theory (DFT) \cite{Hohenberg:1964} simulations using the Vienna \emph{Ab-Initio} Simulation Package (VASP) \cite{Kresse:1996,Kresse:1999,Gajdos:2006,Steiner:2016}.
In all calculations we account for non-collinear magnetization and take spin-orbit coupling (SOC) \cite{Steiner:2016} into account.
The generalized-gradient approximation (GGA) as parametrized by Perdew, Burke, and Ernzerhof (PBE)  \cite{Perdew:1997} is used to describe exchange and correlation.
Kohn-Sham states are expanded into a plane-wave basis with a kinetic-energy cutoff of 600 eV.
A $15\times 15\times 5$ Monkhorst-Pack grid \cite{Monkhorst:1976} was used for  Brillouin zone sampling for all M$_2$As materials.
Convergence tests showed that this leads to an accuracy within 1 meV per 12-atom unit cell for the total energy.
To accelerate self-consistent minimization of the metallic electronic ground state of these materials, we used Gaussian smearing of 25 meV in all calculations.

\begin{figure}
\includegraphics[width=0.95\columnwidth]{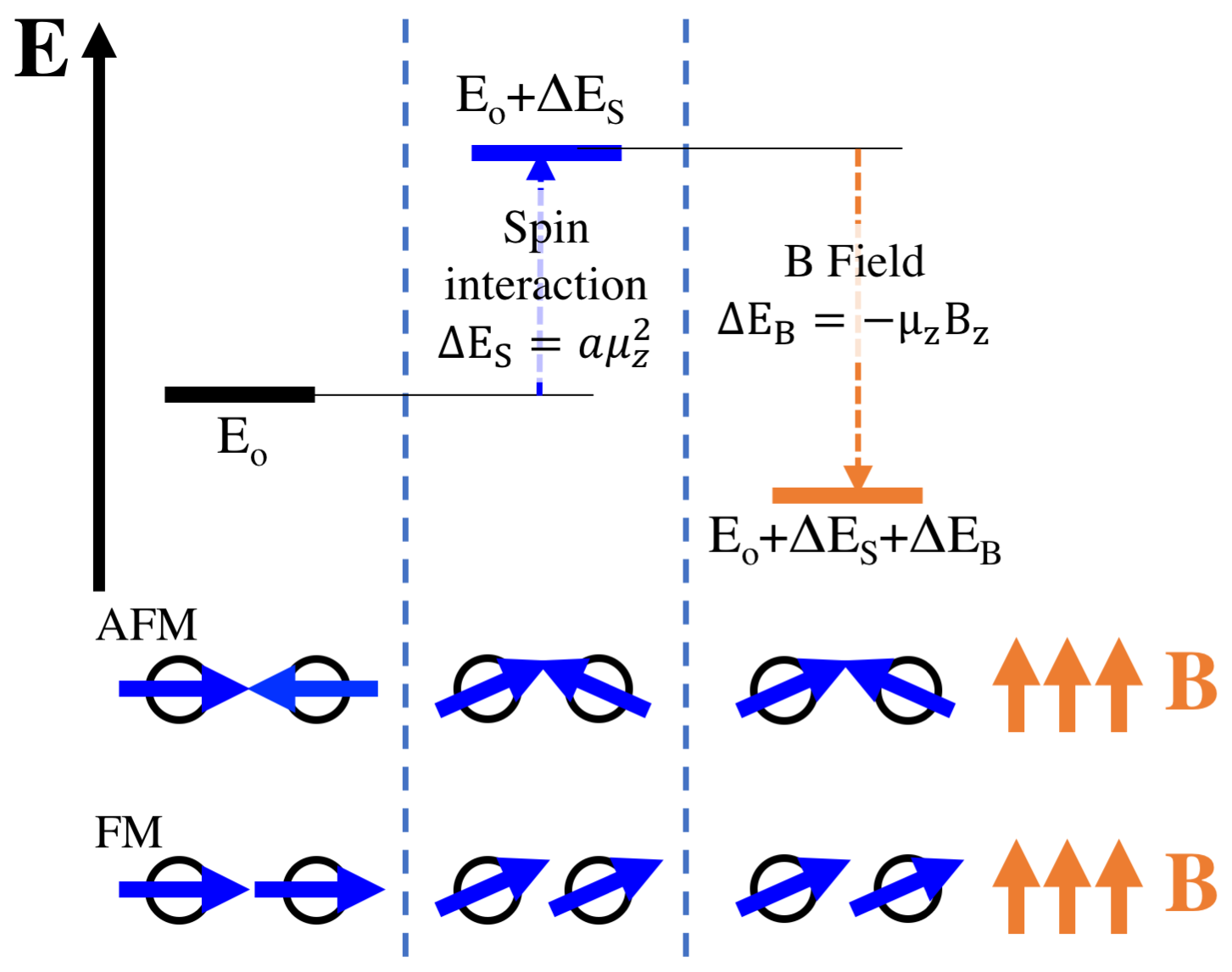}
\caption{\label{fig:M2As-Sus}(Color online.)
Total energy schematic for the AFM ground state $E_0$, the spin-tilted configuration $E_0+\Delta E_\mathrm{S}$, and taking into account an external magnetic field $B$ as the origin of the tilting.
The spin-tilted state has higher energy with respect to the ground state because of spin-spin interaction energy ($\Delta E_{\mathrm{S}}$); an external magnetic field can lower its total energy ($\Delta E_{\mathrm{B}}$).
These energies are a function of sub-lattice magnetization along the $z$-axis ($\mu_{z}$).
Due to the spin-spin interaction, AFM coupling is energetically less favorable under an external magnetic field, compared to FM coupling.
}
\end{figure}

Magnetic moments of a material can tilt perpendicularly to the AF configuration in response to an external magnetic field (see Fig.\ \ref{fig:M2As-Sus}).
In order to implement the magnetic moment tilting, the magnitude of each magnetic moment was fully relaxed but their direction was strongly constrained.
This is achieved in the VASP code by introducing an energy penalty term, so that the total energy becomes
\begin{equation}
\label{eq:constraint}
E=E_0 + \sum_{I}{\lambda\left[\mathbf{M}_I - \mathbf{M}_I^0\left(\mathbf{M}_I^0\cdot\mathbf{M}_I\right)\right]^2}.
\end{equation}
The sum runs over all sites $I$,  $\mathbf{M}_I^0$ is the direction of the constraint at each atomic site, and $\mathbf{M}_I$ is the integrated magnetic moment at site $I$ \cite{VASPmanual:2019}.
We use $\lambda$=50 for the penalty parameter in this work and verified that this choice maintains the imposed moment tilting.
The energy penalty term in this case amounts to about 0.3 $\mu$eV for a tilting of 10$^\circ$.
This tilting also changes the magnetic order and breaks the magnetic symmetry of the material, leading to a net magnetization along the field direction.
For each of these tilted states, we compute lattice parameters and atomic geometries by minimizing Hellman-Feynman forces and we subsequently compute the frequency-dependent complex dielectric tensor \cite{Gajdos:2006}.

When magnetic moments are tilted by the field, FM coupling, as illustrated in Fig.\ \ref{fig:M2As-Sus}, is energetically favorable because the tilting does not break the parallel alignment of the moments, which would affect the exchange interaction.
AF coupling, however, reduces the antiparallel alignment by tilting against the exchange interacting configuration, which is energetically less favorable.
Extremely large magnetic fields can tilt the magnetic moments and switch the material into FM ordering.
Here we study moment-tilted states from $0^{\circ}$ to $10^{\circ}$ in $1^{\circ}$ increments and from $10^{\circ}$ to $90^{\circ}$ in $10^{\circ}$ increments.
We note that for Mn$_2$As and Cr$_2$As a tilting of $1^{\circ}$ corresponds to an external field of about 100 T.

From these DFT simulations we obtain total energies of the AF ground states, $E_0$, as well as total energies of the different moment-tilted configurations, $E_0+\Delta E_\mathrm{S}$ (see Fig.\ \ref{fig:M2As-Sus}).
In addition to computing $E_0+\Delta E_\mathrm{S}$ from DFT, we also model $\Delta E_\mathrm{S}$ as the spin-spin interaction described via an exchange term due to the induced sublattice magnetization, ignoring classical dipole-dipole contributions.
We then account for the energy of a magnetic moment due to net magnetization in an external magnetic field.
This allows us to express the total energy as
\begin{equation}
\label{eq:etot}
E_\mathrm{tot}=E_0+\Delta E_\mathrm{S}+\Delta E_{B}=E_0+a\mu^{2}_{z}-\mu_{z}B_{z},
\end{equation}
where $\mu_{z}$ is the $z$ component of the sub-lattice magnetization, $B_{z}$ is the external magnetic field, and $a$ is a fit parameter (see below).
This leads to our approach for computing the magnetic susceptibility of AF M$_2$As, by plotting the total energy change, $\Delta E_\mathrm{S}$, versus the moment-tilting angle and fit to a quadratic curve.
We use that the magnetic moments of AF M$_2$As in an external magnetic field tilt such that $E_\mathrm{tot}$ is minimal.
The minimization of Eq.\ \eqref{eq:etot} then leads to a relationship between the $z$ component of the sub-lattice magnetization $\mu_{z}$ and the external magnetic field,
\begin{equation}
\label{eq:Mag-sus}
B_{z}=2a\mu_{z}.
\end{equation}
Together with the definition of the magnetic susceptibility
\begin{equation}
\label{eq:maxwell}
B=\mu_{0}(M+H)=\mu_{0}(1+\chi_{v})H,
\end{equation}
we find
\begin{equation}
\label{eq:Mag-sus2}
\chi_{v}=\frac{\mu_{0}}{2a-\mu_{0}},
\end{equation}
where $\mu_{0}$ is the vacuum permeability.

We also use the Kohn-Sham eigenvalues and single-particle states to study the electronic band structure and to compute the complex, frequency-dependent dielectric tensors of M$_2$As, including the anomalous Hall conductivity contribution \cite{Sangalli:2012}, using the VASP code  \cite{Gajdos:2006}.
We note that Drude like intraband contributions in the low-energy range of the frequency-dependent dielectric tensors are not included in our simulations.
From the diagonal and off-diagonal components of the dielectric tensor we compute complex PMOKE signals using \cite{Sangalli:2012}
\begin{equation}
\label{eq:PMOKE}
\Psi_{K}(\omega)=\theta_{K}(\omega)+i\gamma_{K}(\omega)=\frac{-\epsilon_{xy}}{(\epsilon_{xx}-1)\sqrt{\epsilon_{xx}}}.
\end{equation}

\section{\label{sxn:expr}Experimental Methods}

We made single crystals of Fe$_2$As from the melt\cite{Yang2019}.
The Fe$_2$As sample was polished along the (001) orientation with an Allied Multiprep automatic polisher with diamond lapping films down to 0.3 $\mu$m.
The orientation was observed via X-ray diffraction pole figures.
The miscut of the surfaces is within 10$^{\circ}$.
Right after polish, about 5 nm Al$_2$O$_3$ was deposited with Atomic Layer Deposition (ALD) to prevent oxidation of the surface.

Our polar MOKE measurement setup was described in detail in Ref.\ \onlinecite{kimling2017thermal}.
We used a Ti:Sapphire laser with 80 MHz repetition rate and 783 nm center wavelength as light source.
The probe beam was modulated at 200 Hz with a chopper and was used to do the measurement.
A balanced detector connected with an AF lock-in amplifier was used to measure the signal.
The external field was applied by an electromagnet.
There is a hole for one of the poles so that light can pass through and both light and field are perpendicular to the sample surface.
The gap between both poles is within 1 cm.
The magnetic field was controlled with a power source and measured with a Gauss meter. 

The polar MOKE signal at field $H$ is \cite{yang2020thesis}
\begin{equation}
\theta = \frac{I_+-I_-}{2(I_++I_-)},
\end{equation}
where $I_+$ and $I_-$ are the light intensities for $+H$ and $-H$ fields, respectively.

\section{\label{sxn:results}Results and Discussion}

\subsection{\label{sxn:gsst}Atomic geometries}

\begin{figure}
\includegraphics[width=0.98\columnwidth]{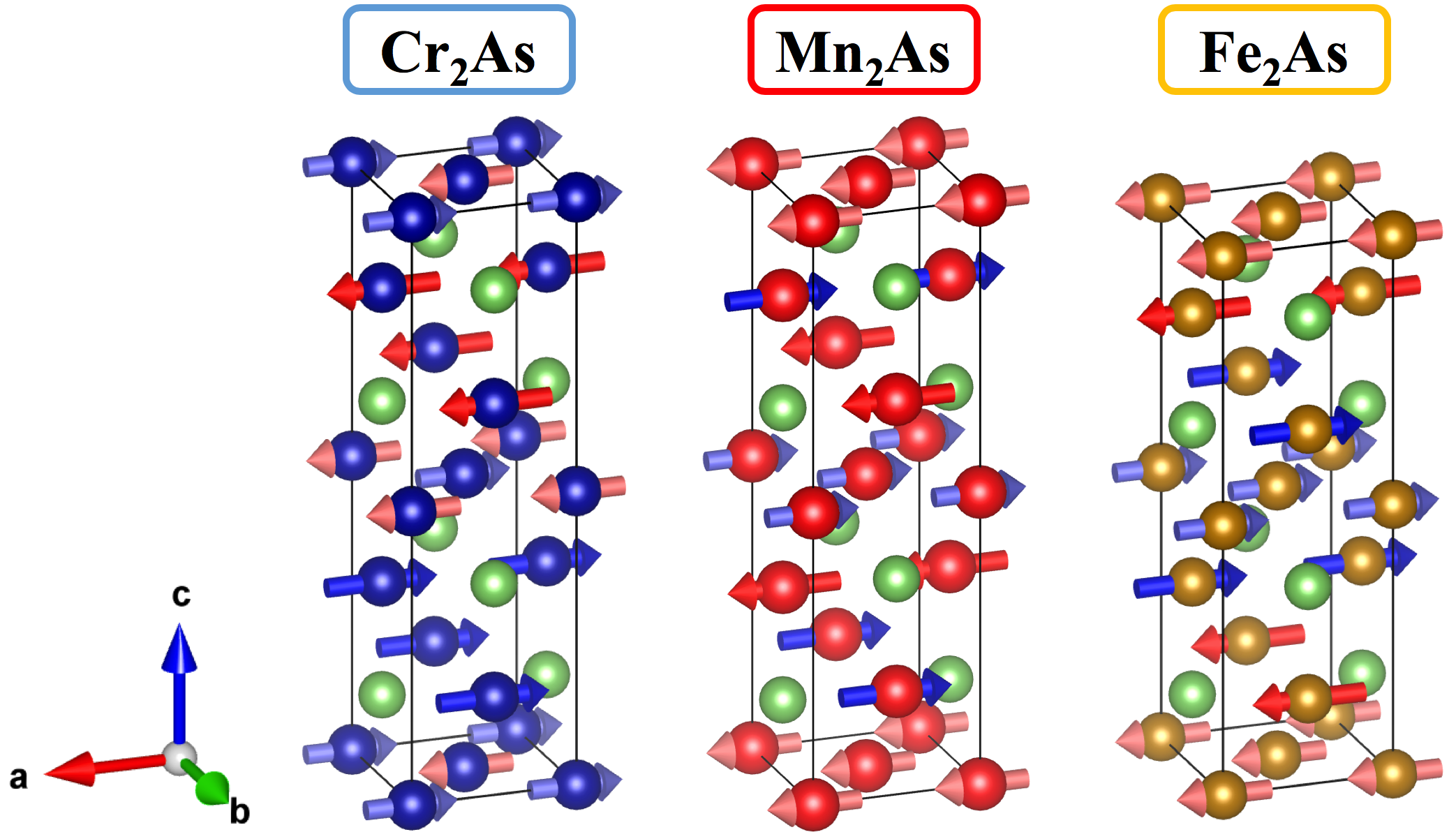}
\caption{\label{fig:M2As-str}(Color online.)
Chemical and magnetic structure of Cr$_2$As, Mn$_2$As, and Fe$_2$As shown for the magnetic primitive unit cells.
Chromium atoms are blue, manganese are red, iron are gold, and arsenic are green.
Magnetic cations of type I and II, M(I) and M(II), are chemically and magnetically non-equivalent and are indicated by light and dark color of blue and red arrows, respectively.
The color of the arrows (blue vs.\ red) illustrates their anti-parallel orientation.
}
\end{figure}

\begin{table}
\caption{\label{tab:latmag}
Lattice parameters (in \AA) and (untilted) magnetic moments (in $\mu_\mathrm{B}$) for Cr$_2$As, Mn$_2$As, and Fe$_2$As, compared to experimental values.
Relative deviation from experimental values is given as $\Delta_\mathrm{exp}$.
All experimental values are measured at room temperature, while the magnetic moments of Fe$_2$As in parentheses are extrapolated to 0 K, based on temperature-dependent measurements.
Tilting of magnetic moments by $10^{\circ}$ influences the lattice parameters by less than 0.1\,\%.
}
\begin{tabular}{cccccc}
\hline
Cr$_2$As& $a$ & $b$ & $c$ & $\mu_\mathrm{Cr(I)}$ & $\mu_\mathrm{Cr(II)}$ \\
\hline
This work & 3.56&3.59&12.63&1.11&1.98\\
Exp.\ \cite{Yamaguchi:1972} &3.60&3.60&12.68&0.4&1.34\\
$\Delta_\mathrm{exp}\,(\%)$&$-0.8$&0.1&$-0.5$&178&47.8\\
\hline
Mn$_2$As& $a$ & $b$ & $c$ & $\mu_\mathrm{Mn(I)}$ & $\mu_\mathrm{Mn(II)}$ \\
\hline
This work&3.68&3.68&12.27&1.87&3.24\\
Exp.\ \cite{Austin:1962} &3.78&3.78&12.56&3.70&3.50\\
$\Delta_\mathrm{exp}$\,(\%) & $-2.7$ & $-2.7$ & $-2.3$ & $-49.5$ & $-7.4$\\
\hline
Fe$_2$As& $a$ & $b$ & $c$ & $\mu_\mathrm{Fe(I)}$ & $\mu_\mathrm{Fe(II)}$\\
\hline
This work & 3.624 & 3.624 & 11.72 & 1.23 & 2.25\\
\multirow{2}{*}{Exp.\ \cite{Katsuraki:1966} } &\multirow{2}{*}{3.627}&\multirow{2}{*}{3.627}&\multirow{2}{*}{11.96}&0.95&1.52\\
&   &   &   &(1.28) &(2.05)\\ 
\multirow{2}{*}{$\Delta_\mathrm{exp}$\,(\%)} & \multirow{2}{*}{$-0.1$} & \multirow{2}{*}{$-0.1$} & \multirow{2}{*}{$-2.0$} & $29.5$ & 95.3\\
&   &   &   &($-3.9$) &(9.8)\\ 
\hline
\end{tabular}
\end{table}

The atomic and magnetic structure of Cr$_2$As, Mn$_2$As, and Fe$_2$As in the magnetic ground state is shown in Fig.\ \ref{fig:M2As-str}.
Lattice parameters and magnetic moments, computed from fully relaxed atomic geometries using DFT, are listed in Table \ref{tab:latmag}.
The chemical structure and space group is $P_{4}/nmm$ for all three materials.
While the magnetic ordering differs for these three M$_2$As compounds, all magnetic space and point groups are the same, $P_{a}nma$ and $mmm1'$, respectively.
Tilting the magnetic moments changes the magnetic structure and breaks this symmetry:
The point groups become $Pnm'a$ and $m'm'm$.
We identified all chemical and magnetic symmetries using FINDSYM in the ISOTROPY Software Suite \cite{Stokes:2005, FINDSYM}.
We compare all symmetry operations and find that tilting of the magnetic moments breaks time reversal symmetry ($T$: +1 $\rightarrow$ $-1$), thus, enabling these M$_2$As materials to exhibit PMOKE signals.

Comparison to literature data shows that our results for lattice parameters agree to within 3\,\% or better with experiment\cite{Yamaguchi:1972, Austin:1962, Katsuraki:1966}.
The large discrepancy for the amplitude of the magnetic moments of Cr$_2$As and Fe$_2$As (see Table \ref{tab:latmag}) can be explained by the experiments being carried out at room temperature, i.e., close to the corresponding N\'eel temperatures.
Katsuraki \emph{et al.}\ \cite{Katsuraki:1966} extrapolate the magnetic moment to 0 K based on the magnetic intensity from neutron scattering, which shows agreement within 10\,\% with our simulations.
For Mn$_2$As, the measured magnetic moment is larger than our calculated results.
Similar calculations for Mn$_2$As are reported by Zhang \emph{et al.}\ and point out that this deviation originates from the lack of on-site Coulomb interaction of localized electrons \cite{Zhang:2013}.

\subsection{\label{sxn:ms}Magnetic susceptibilities}

\begin{figure}
\includegraphics[width=0.99\columnwidth]{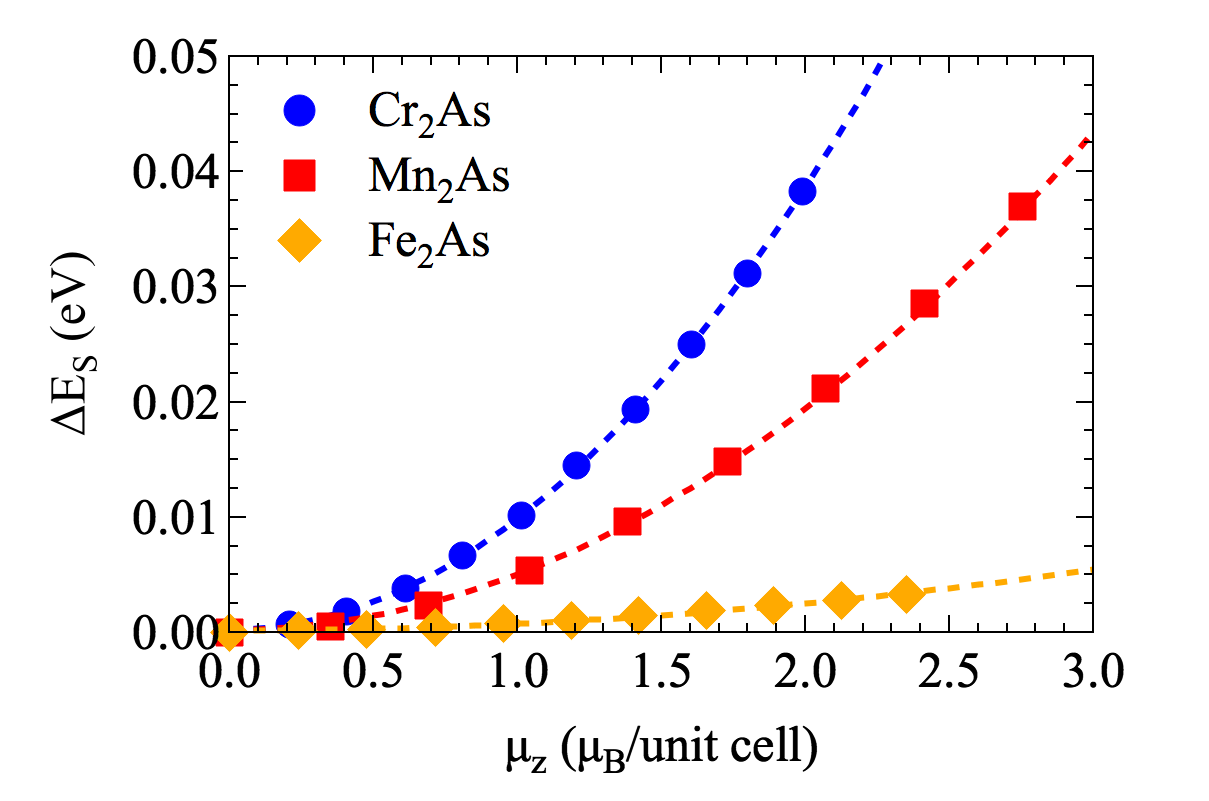}
\caption{\label{fig:M2As-eng}(Color online.)
Each dot represents the relative energy change $\Delta E_\mathrm{S}$ with respect to $E_0$ for moment-tilted states of Cr$_2$As, Mn$_2$As, and Fe$_2$As, see Fig.\ \ref{fig:M2As-Sus}.
Neighboring data points correspond to a tilting difference of $1^{\circ}$.
Dashed lines show quadratic fits to the data.
}
\end{figure}
   
The magnetic susceptibility is important to understand the response of the magnetic structure of antiferromagnetic materials (AFMs) to external magnetic fields.
It determines the orientation change of the magnetic moments in the material for a given applied external magnetic field, as discussed in Sec.\ \ref{sxn:comp}.
Magnetic susceptibility measurements were performed by Yuzuri \emph{et al.}\ on polycrystalline Cr$_2$As\cite{Yuzuri:1960} and Mn$_2$As\cite{Yuzuri2:1960}.
The measured magnetic susceptibility of polycrystalline materials corresponds to a directional average.
The polycrystalline measurements report that the unitless magnetic susceptibility of Cr$_{2.1}$As \cite{Yuzuri:1960} is $9.67 \times 10^{-4}$ at $273$\,K and of Mn$_{2.3}$As \cite{Yuzuri2:1960} is $1.84 \times 10^{-3}$ at $465$\,K.
For Fe$_2$As, the susceptibility was measured along the two different directions $a$ ($\chi_{a}$) and $c$ ($\chi_{c}$), see Fig.\ \ref{fig:M2As-str}, of a single-crystalline sample \cite{Yang:2020},
corresponding to the [100] and [001] directions \cite{Yang:2020}.
At 4\,K temperature, the magnetic susceptibility along the $a$ and $c$ axes is $0.011$ and $0.015$, respectively.

In Fig.\ \ref{fig:M2As-eng} we illustrate the increase in total energy $\Delta E_s$ when tilting the moment, leading to a non-vanishing magnetization along the $z$ axis.
From the parabolic fits in this figure we determine the curvature $a$ in Eq.\ \eqref{eq:etot}, and calculate the corresponding magnetic susceptibilities using Eq.\ \eqref{eq:Mag-sus2}.
Our DFT results show that the magnetic susceptibility perpendicular to N\'eel vector along $c$-axis of Cr$_2$As, Mn$_2$As, and Fe$_2$As is $2.10 \times 10^{-4}$, $4.14 \times 10^{-4}$, and $3.55 \times 10^{-3}$, respectively.
The direct comparison between computational and measured results is impractical since the definition of magnetic susceptibility is different, and the susceptibility change from thermal contribution is non-negligible.

We also note that the energy change for a given $z$ axis magnetization is much smaller for Fe$_2$As, compared to the other two materials (see Fig.\ \ref{fig:M2As-eng}).
The difference in magnitude of the magnetic susceptibilities of M$_2$As can be explained by their magnetic structures (see Fig.\ \ref{fig:M2As-str}).
Fe$_2$As has a unique magnetic structure, in that each chemical unit cell is ferromagnetically ordered and the cells are layered with alternating antiparallel moment directions.
This synthetic AF-like structure allows for easy tilting of the moments by an external magnetic field.
However, Cr$_2$As and Mn$_2$As do not have this feature, which leads to smaller magnetic susceptibility.
This is consistent with effective exchange parameters, e.g.\ computed by Zhang \textit{et al.}\cite{Zhang:2013} 
They report a negative sign for first- and second-nearest neighbor atoms in Cr$_2$As and Mn$_2$As, to explain that these materials are antiferromagnetically coupled.
Conversely, Fe$_2$As is more ferromagnetically coupled with only one antiferromagnetic coupling constant \cite{Zhang:2013}.
Figure \ref{fig:M2As-Sus} illustrates that under an external magnetic field the ferromagnetically coupled moments change their orientation, while keeping their parallel alignment.
This orientation change only needs to overcome the anisotropy energy.
Conversely, antiferromagnetically coupled moments break their antiparallel alignment upon tilting, leading to an exchange interaction, see Fig.\ \ref{fig:M2As-Sus}.
This explains a larger magnetic susceptibility of more ferromagnetically coupled Fe$_2$As, compared to the the smaller magnetic susceptibility of Mn$_2$As and Cr$_2$As.
This also means that Fe$_2$As generates larger net magnetization along the field direction under a given external magnetic field and we revisit this point later in our discussion of PMOKE for different field strengths.

\subsection{\label{sxn:bsa}Band structure analysis}

\begin{figure}
\includegraphics[width=0.99\columnwidth]{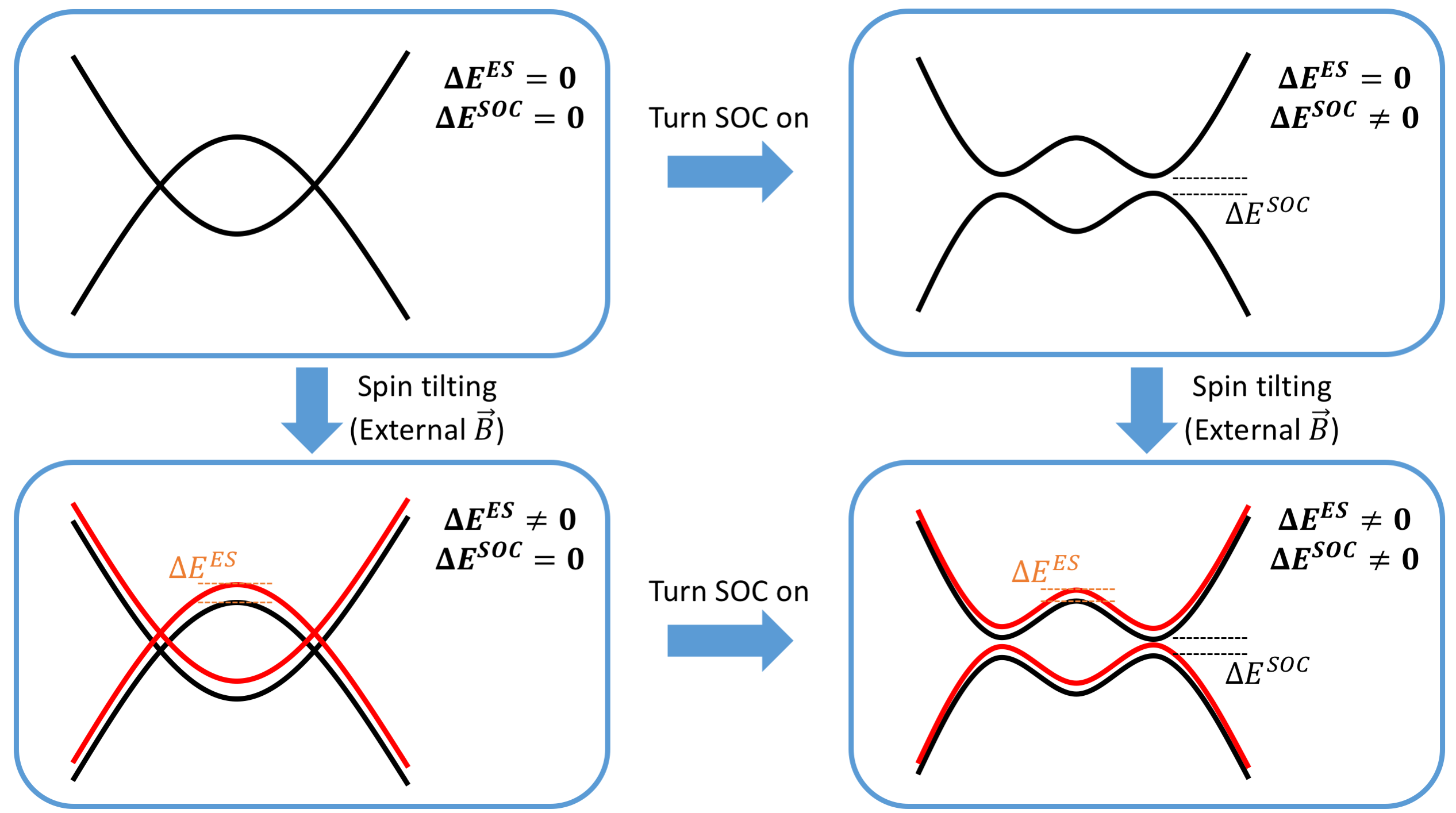}
\caption{\label{fig:M2As-band}(Color online.)
Schematic illustration of the effects of exchange ($\Delta E^\mathrm{ES}$) and spin-orbit splitting ($\Delta E^\mathrm{SOC}$) on the ground-state electronic band structure.
The effect of exchange splitting is due to the magnetic moment tilting.
}
\end{figure}

To understand how a PMOKE signal arises in antiferromagnetic materials under an external magnetic field, we first explore the effect of such a field on the electronic band structure.
For ferromagnetic materials, Oppeneer \textit{et al.}\ discuss that essential parameters of PMOKE are spin-orbit coupling, exchange splitting, and (strain dependent) lattice spacing \cite{Oppeneer:1992}.
The lattice relaxation due to moment-tilting is less than 0.1\% and, therefore, negligible.
Hence, we focus on spin-orbit coupling and exchange splitting, as illustrated in Fig.\ \ref{fig:M2As-band}.

To this end, we define a measure $\Delta \bar{E}^\mathrm{SOC}$ for the spin-orbit splitting,
\begin{equation}
\label{eq:SOC}
\Delta \bar{E}^\mathrm{SOC}=\sum_{\mathbf{k},i} \frac{|\varepsilon^\mathrm{SOC}(\mathbf{k},i)-\varepsilon^\mathrm{no SOC}(\mathbf{k},i)|}{N_{k}N_B}.
\end{equation}
Here, $\mathbf{k}$ indexes all $N_{k}$ points in the Brillouin zone, and $i$ is the band index running over all $N_B$ bands.
We find that the dependence on $N_B$ is weak (see the supplemental material).
Kohn-Sham energies $\varepsilon^\mathrm{SOC}(\mathbf{k},i)$ and $\varepsilon^\mathrm{no SOC}(\mathbf{k},i)$ result from non-collinear density functional theory simulations with and without inclusion of spin-orbit coupling, respectively \cite{Steiner:2016}, corresponding to the difference of the left and right panels of Fig.\ \ref{fig:M2As-band}.
Actual band-structure data is shown in the supplemental material.
Similarly, we quantify the effect of exchange splitting using \begin{equation}
\label{eq:ES}
\Delta \bar{E}^\mathrm{ES}=\sum_{\mathbf{k},i} \frac{|\varepsilon^\mathrm{maj}(\mathbf{k},i)-\varepsilon^\mathrm{min}(\mathbf{k},i)|}{N_k N_B},
\end{equation}
where $\varepsilon^\mathrm{maj}(\mathbf{k},i)$ and $\varepsilon^\mathrm{min}(\mathbf{k},i)$ are majority and minority spin Kohn-Sham energies of the $i^{th}$ band at $\mathbf{k}$.
While in the ground state all bands of antiferromagnetic materials are degenerate, moment-tilting leads to a net magnetization in field direction and induces a splitting (see difference between top and bottom panels of Fig.\ \ref{fig:M2As-band}).

\begin{figure}
\includegraphics[width=0.98\columnwidth]{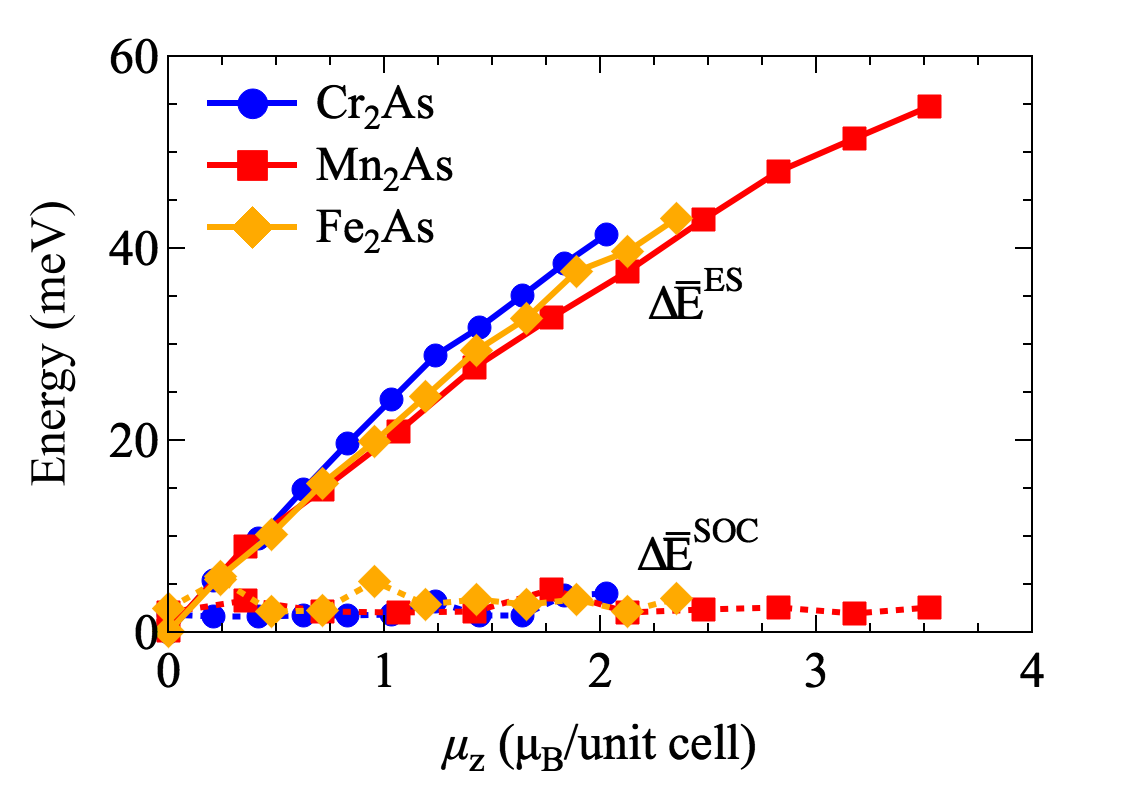}
\caption{\label{fig:M2As-energy}(Color online.)
Energy splittings $\Delta \bar{E}^\mathrm{ES}$, Eq.\ \eqref{eq:ES}, and $\Delta \bar{E}^\mathrm{SOC}$, Eq.\ \eqref{eq:SOC}, vs.\ magnetization for different tilting of magnetic moments.
Exchange splitting increases with magnetization, while spin-orbit coupling remains almost constant.
Thus, we attribute PMOKE in antiferromagnetic M$_2$As under external magnetic fields to exchange splitting.
}
\end{figure}

The resulting energy splittings $\Delta \bar{E}^\mathrm{ES}$ and $\Delta \bar{E}^\mathrm{SOC}$ are shown as a function of the magnetic moment tilting between $0^{\circ}$ and $10^{\circ}$ in Fig.\,\ref{fig:M2As-energy}.
This illustrates that while increased tilting leads to increased net magnetization, spin-orbit coupling is almost unaffected.
At the same time, exchange-splitting gradually increases for all three antiferromagnetic M$_2$As materials.
Thus, we conclude that the origin of PMOKE under external magnetic fields is purely due to the change of exchange splitting with tilting.
We also note that the electronic band structure exhibits small spin-orbit induced splitting even in the (untilted) antiferromagnetic ground state.
Hence, the vanishing PMOKE signal in the ground state is also explained exclusively by the absence of exchange splitting in this case.
This interpretation is consistent with symmetry analysis which showed that non-zero PMOKE occurs when the magnetic symmetry is broken by a non-zero net magnetization.

\subsection{\label{sxn:pmoke}Polar magneto-optical Kerr effect}

Tilting of magnetic moments affects the dielectric tensor, which directly determines the linear magneto-optical Kerr tensor \cite{Eremenko:1992}.
In particular, the $mmm1'$ point group does not allow for linear MOKE by symmetry, while the dielectric tensor of the $m'm'm$ point group has off-diagonal elements that cause linear MOKE via Eq.\ \eqref{eq:PMOKE}.
We then use Eq.\,\eqref{eq:PMOKE} to compute PMOKE rotation $\theta_K$ and ellipticity $\gamma_K$ from the complex frequency-dependent dielectric tensor of antiferromagnetic M$_2$As under different magnetic moment tilting angles.
In addition, the magnitude of the PMOKE response depends on the magnitude of the magnetic moment tilting.
We distinguish two angle ranges:
For low tilting angles, approximately between 0$^{\circ}$ and 10$^{\circ}$, PMOKE depends linearly on the external field.
Conversely, for tilting angles higher than that (up to 90$^{\circ}$), the response becomes non-linear.
Using our computed magnetic susceptibility, magnetic moment tilting can be connected to the strength of the external magnetic field.

\begin{figure}
\includegraphics[width=0.98\columnwidth]{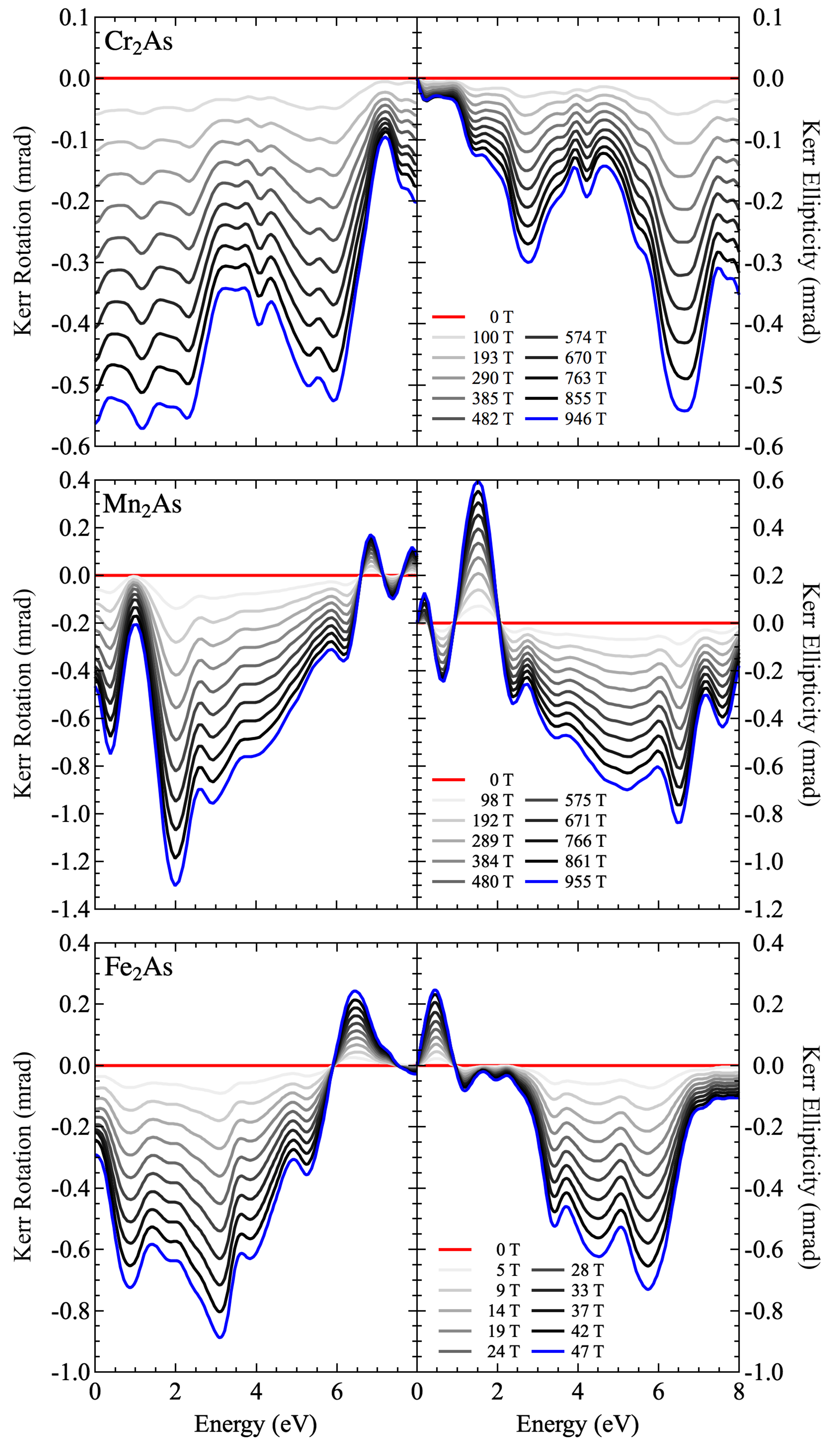}
\caption{\label{fig:Kerr}(Color online.)
PMOKE rotation $\theta_K$ and ellipticity $\gamma_K$ for antiferromagnetic Cr$_2$As, Mn$_2$As, and Fe$_2$As under external magnetic fields corresponding to moment-tilting angles from 0$^{\circ}$ to 10$^{\circ}$ in 1$^{\circ}$ increments.
The magnetic field values are computed using the magnetic susceptibility from our DFT results).
}
\end{figure}

For the low angle range, Fig.\ \ref{fig:Kerr} illustrates a gradual increase of the PMOKE response as the external magnetic field increases.
Peak positions and overall spectral shape are unaffected by the field magnitude.
The majority of experimental MOKE studies is done in the visible spectral range.
Thus, from our data we find the wavelength that maximizes the MOKE signal in this range.
The highest peak of the Kerr rotation spectra occurs at 1.13 eV (Cr$_2$As), 1.99 eV (Mn$_2$As), and 3.08 eV (Fe$_2$As).
In the case of Cr$_2$As this peak is outside the visible spectral range between 1.66 eV and 3.30 eV and the highest peak within the visible spectrum occurs at 2.27 eV.
For Kerr ellipticity, the maxima occur at 6.65 eV (Cr$_2$As), 6.46 eV (Mn$_2$As), and 5.75 eV (Fe$_2$As) and the highest values in the visible spectral range are at 2.76 eV, 1.66 eV, and 3.30 eV, respectively.

\begin{figure}
\includegraphics[width=0.8\columnwidth]{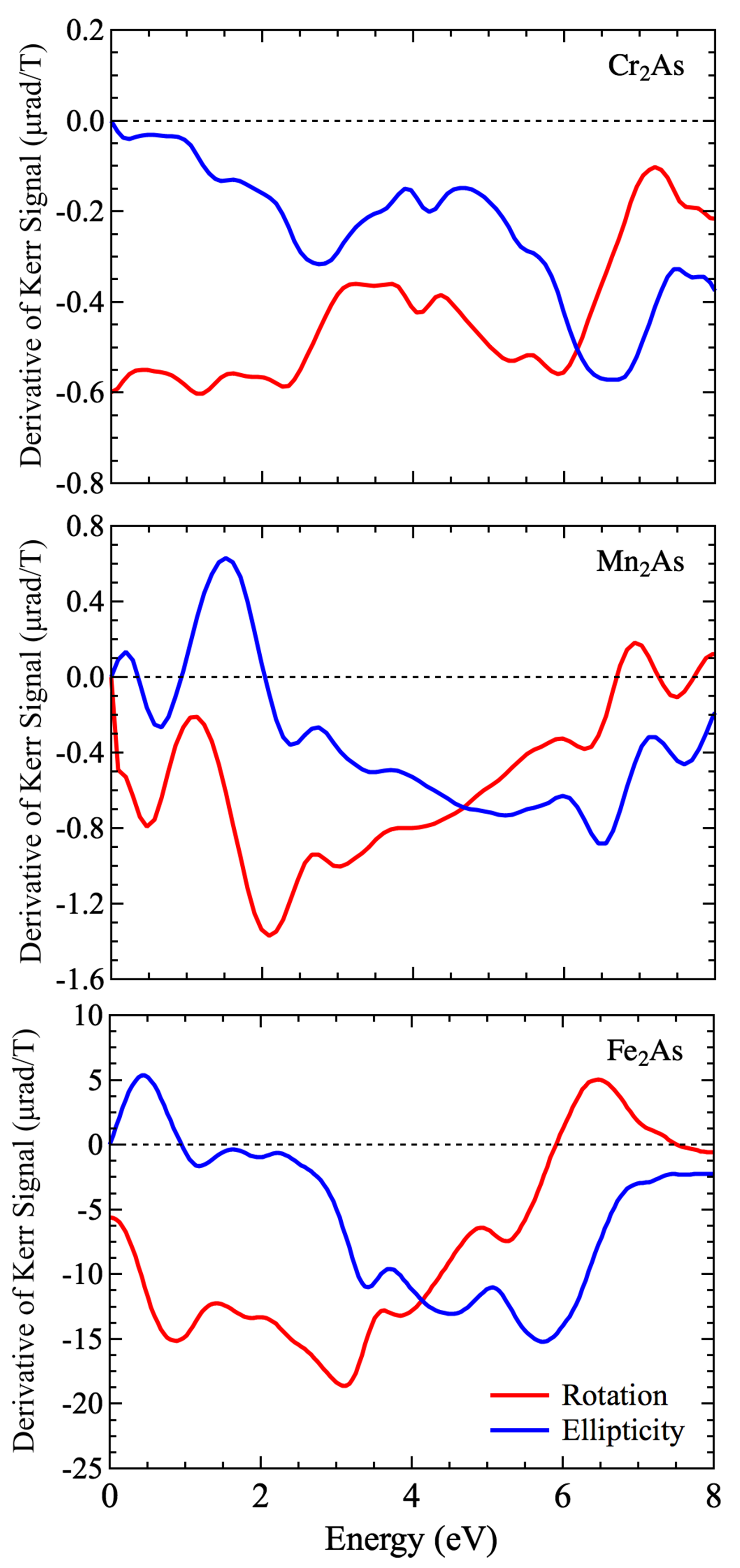}
\caption{\label{fig:M2As-Kerr-gradient}(Color online.)
PMOKE signal gradient spectra, $d\theta_K/dB$ (red solid lines) and $d\gamma_K/dB$ (blue solid lines), of antiferromagnetic M$_2$As under an external magnetic field of 1 T.
}
\end{figure}

Due to the small magnetic susceptibilities of Cr$_2$As, Mn$_2$As, and Fe$_2$As moment-tilting of only 1$^{\circ}$ still corresponds to large external magnetic fields.
At the same time, implementing small tilting angles using the constraint in Eq.\ \eqref{eq:constraint} poses numerical challenges.
Hence, we use that the MOKE spectra depend linearly on the magnetic field for small tilting (see Fig.\ \ref{fig:Kerr}) and linearly interpolate the response into the range of magnetic fields of practical importance.
To this end, we compute the gradients $d\theta_K/dB$ and $d\gamma_K/dB$ for each photon energy, using the PMOKE spectra for tilted moments between 0$^\circ$ and 10$^\circ$, and the constraint of vanishing PMOKE for 0$^\circ$ tilting.
We then use these gradients to compute the PMOKE spectra for all three materials under an external magnetic field of 1 T, see Fig.\ \ref{fig:M2As-Kerr-gradient}.
From this we find that while the PMOKE spectra exhibit a significant dependence on the photon energy, their overall magnitude is strongly influenced by the size of magnetic susceptibility (Cr$_2$As<Mn$_2$As<Fe$_2$As).
In particular, the larger magnitude of the gradient observed for Fe$_2$As (see Fig.\ \ref{fig:M2As-Kerr-gradient}), compared to Cr$_2$As and Mn$_2$As, is due to the larger magnetic susceptibility of Fe$_2$As.
Therefore, besides spin-orbit coupling and exchange splitting effect, magnetic susceptibility becomes another essential parameter to understand PMOKE from antiferromagnetic materials under external magnetic field.

\begin{figure}
\includegraphics[width=0.98\columnwidth]{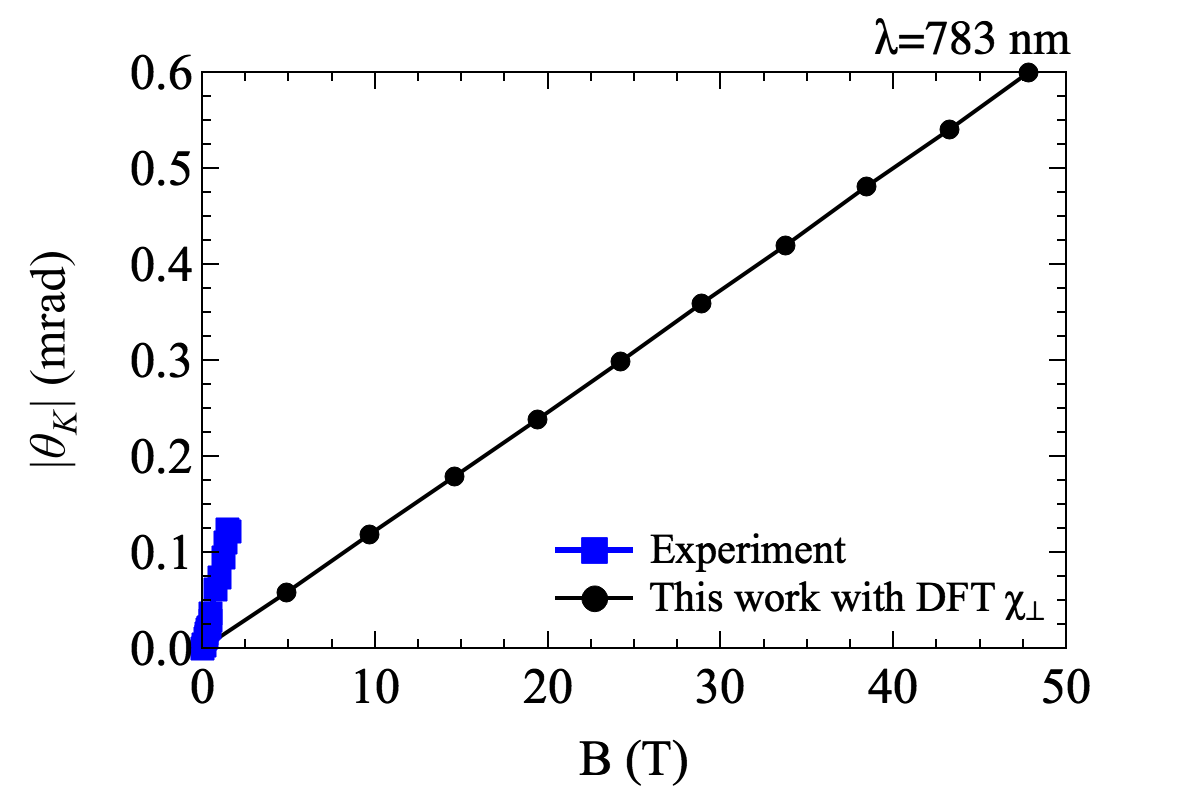}
\caption{\label{fig:Fe2As-Kerr-linear}(Color online.)
The absolute PMOKE rotation $\left|\theta_K\right|$ of antiferromagnetic Fe$_2$As depends linearly on moment tilting between 0$^{\circ}$ and 10$^{\circ}$ at a wavelength of 783 nm.
The magnetic field is computed from the tilting using different susceptibilities:
The black line uses the DFT results from this work, while red and orange lines are computed using experimental perpendicular magnetic susceptibility ($\chi_{\bot}$) and the magnetic susceptibility along the $c$ axis ($\chi_{c}$) in Ref.\ \onlinecite{Yang:2020}, respectively.
Blue line represents our experiment.
}
\end{figure}

In addition, in Fig.\ \ref{fig:Fe2As-Kerr-linear} we show the linear response of the Kerr rotation at a wavelength of 783 nm for different tilting angles.
We compute the corresponding external magnetic field $\mathbf{H}$ from the net magnetization $\mathbf{M}$ using the magnetic susceptibility via $\mathbf{M}=\chi\mathbf{H}$).
The measured Kerr rotation in Fe$_2$As at an excitation wavelength of 783 nm and an external magnetic field of 1 T is about 74.1 $\mu$rad.
When using the DFT susceptibility we underestimate the Kerr rotation signal and find 12.5 $\mu$rad at 1 T.
Experiment and simulation agree that the trend is linear for small external fields, however, the magnitude of the calculated signal deviates from the measured result.

\begin{figure}
\includegraphics[width=0.98\columnwidth]{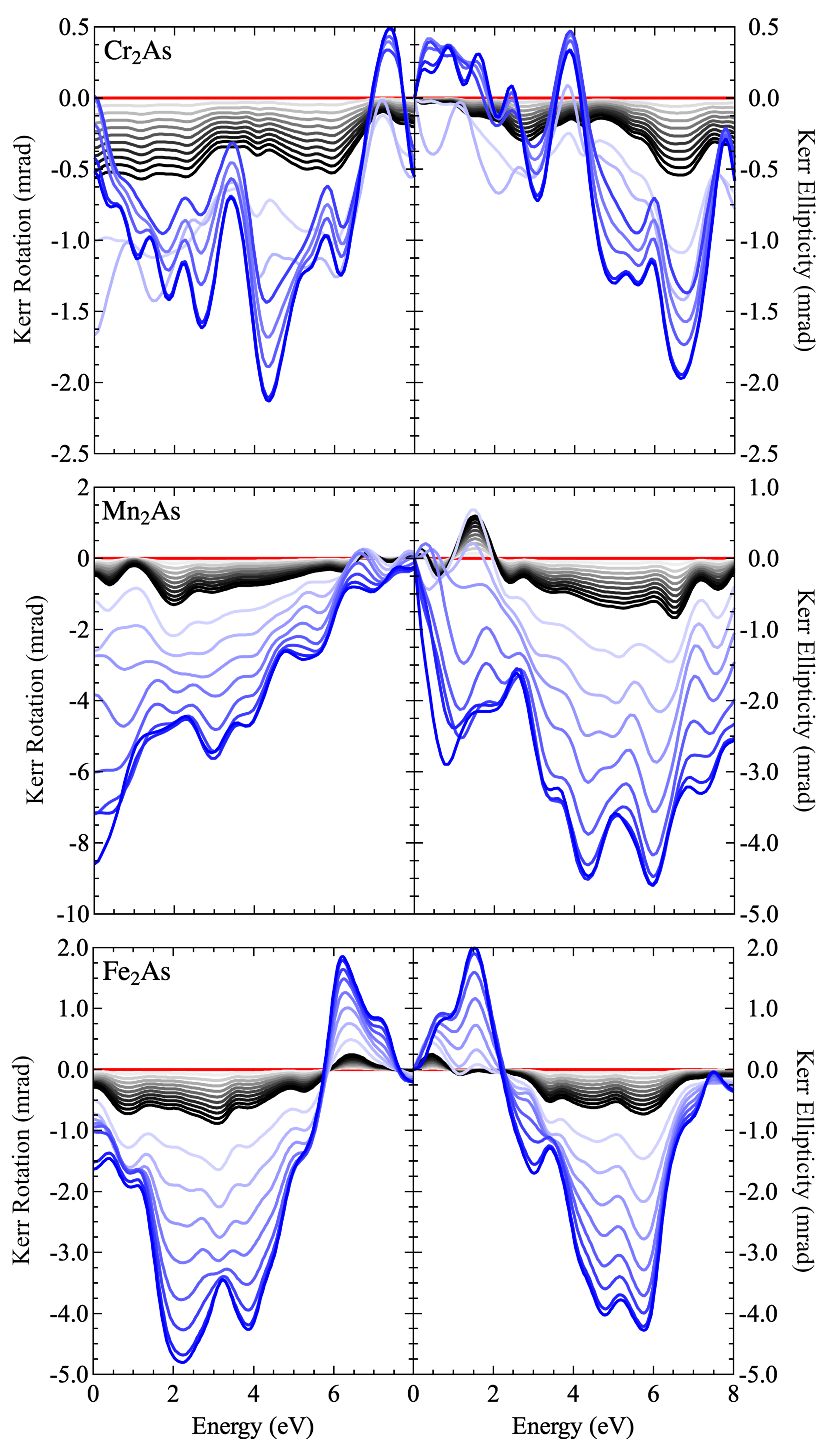}
\caption{\label{fig:Kerr-high}
(Color online.)
PMOKE rotation $\theta_K$ and ellipticity $\gamma_K$ for antiferromagnetic Cr$_2$As, Mn$_2$As, and Fe$_2$As for different moment tilting.
Three different color schemes represent ground state (red line), low tilting angles from 1$^{\circ}$ to 10$^{\circ}$  in 1$^{\circ}$ increments (grey lines), and high tilting angles from 20$^{\circ}$ to 90$^{\circ}$ in 10$^{\circ}$ increments (blue lines).
The low angle range represents linear change of the PMOKE signal and the high angle range illustrates non-linear behavior.
}
\end{figure}

When the tilting angle of the magnetic moments exceeds approximately $10^{\circ}$ the linear response discussed before (see Fig.\ \ref{fig:Kerr}) disappears.
Instead, Fig.\ \ref{fig:Kerr-high} illustrates that in this regime the PMOKE signal displays a complicated dependence on the external magnetic field.
Especially the spectra of Cr$_2$As for tilting angles beyond 10$^{\circ}$ (blue curves) look completely different than the low-angle spectra and the ellipticity near photon energies around 1\,eV even changes its sign.
By analyzing our data for the electronic band structures of these materials (see the supplemental material), we attribute this to a large exchange splitting effect.
Since spin-orbit coupling is simultaneously present, new band crossings and complex splittings are induced, causing the non-linear response of PMOKE signals in this case.
However, this non-linear behavior of PMOKE in M$_2$As (M=Cr, Mn, Fe) requires extremely large external magnetic fields that cannot be reached in experiment.
There are almost no new band crossings for low tilting angles, due to the much smaller exchange splitting.
In this regime, the smaller exchange splitting proportional to the external field directly causes the linear response of PMOKE signals.
Thus, our results demonstrate that PMOKE in antiferromagnetic materials requires a change of the exchange splitting, similar to ferromagnetic materials \cite{Oppeneer:1992}.

\section{\label{sxn:cncl}Conclusion}

Using first-principles electronic-structure calculations based on density functional theory, we predicted the polar magneto-optical Kerr effect spectra for antiferromagnetic M$_2$As (M=Cr, Mn, Fe).
Breaking of the magnetic symmetry is necessary for this effect to appear and we achieve this in this work via external magnetic fields that we simulate using a constraint that tilts the magnetic moments.
We devise a computational scheme to compute magnetic susceptibility from the total energy change upon tilting and find reasonable agreement with experimental results.
Subsequently, we compute the frequency-dependent magneto-optical Kerr effect from the off-diagonal components of the dielectric tensor and study the dependence on the external field.
Our simulation results and experimental data show that the strength of Kerr rotation and ellipticity scales linearly with the field for small fields.
Using our band-structure results we trace this back to exchange splitting and we show that the dependence of spin-orbit on the external field is negligible.
From our results we conclude that spin-orbit coupling, exchange splitting, and magnetic susceptibility are three key parameters that jointly determine the magneto-optical Kerr effect in antiferromagnetic materials under external magnetic fields.

\begin{acknowledgments}
This material is based upon work supported by the National Science Foundation under Grant No.\ DMR-1720633.
This research is part of the Blue Waters sustained-petascale computing project, which is supported by the National Science Foundation (awards OCI-0725070 and ACI-1238993) and the state of Illinois.
Blue Waters is a joint effort of the University of Illinois at Urbana-Champaign and its National Center for Supercomputing Applications.
This work made use of the Illinois Campus Cluster, a computing resource that is operated by the Illinois Campus Cluster Program (ICCP) in conjunction with the National Center for Supercomputing Applications (NCSA) and which is supported by funds from the University of Illinois at Urbana-Champaign.
\end{acknowledgments}

\bibliographystyle{apsrev}
\bibliography{literature.bbl}

\end{document}